\documentclass[journal]{IEEEtran}

\usepackage{amsmath,amssymb}
\usepackage[final]{graphicx}
\usepackage{cite}
\usepackage{subfigure}
\usepackage{comment,color}
\usepackage{setspace}
\usepackage{booktabs}
\usepackage{subfigmat}

\newcommand{\R}{\ensuremath{{\mathbb R}}}

\newcommand{\NN}{{\mathcal N}}

\newcommand{\ra}{\rightarrow}

\newcommand{\zz}{\mathsf{z}}
\newcommand{\sat}{\mathrm{sat}}

\newcommand{\id}{\mathrm{id}}
\newcommand{\XX}{\mathcal X}
\newcommand{\UU}{\mathcal U}

\newcommand{\TT}{{\mathcal T}}

\newtheorem{thm1}{\bf Theorem}
\newtheorem{prop1}{\bf Proposition}
\newtheorem{lem1}{\bf Lemma}
\newtheorem{assmpt1}{\bf Assumption}
\newtheorem{defn1}{\bf Definition}
\newtheorem{rem1}{\bf Remark}
\newtheorem{cor1}{\bf Corollary}

\newenvironment{defn}{\begin{defn1}}{\hfill$\Diamond$\end{defn1}}
\newenvironment{asm}{\begin{assmpt1}}{\hfill$\Diamond$\end{assmpt1}}
\newenvironment{rem}{\begin{rem1}}{\hfill$\Diamond$\end{rem1}}
\newenvironment{lem}{\begin{lem1}}{\hfill$\Diamond$\end{lem1}}
\newenvironment{thm}{\begin{thm1}}{\hfill$\Diamond$\end{thm1}}
\newenvironment{cor}{\begin{cor1}}{\hfill$\Diamond$\end{cor1}}
\newenvironment{prop}{\begin{prop1}}{\hfill$\Diamond$\end{prop1}}

\begin{document}
%
\title{Detection of Sensor Attack and Resilient State Estimation
	for Uniformly Observable Nonlinear Systems having Redundant Sensors}
%
%
%

\author{Junsoo Kim, Chanhwa Lee, Hyungbo Shim, Yongsoon Eun, and Jin H.~Seo
\thanks{This work was supported by the National Research Foundation of Korea (NRF) grant funded by the Korea government (MSIT) (No.~2015R1A2A2A01003878) and by IITP grant funded by the Korea government
(No.~B0101-15-0557, Resilient Cyber-Physical Systems Research).}
\thanks{J.~Kim, C.~Lee, H.~Shim, and J.~H.~Seo are with ASRI, Department of Electrical and Computer Engineering, Seoul National University, Korea.
Y.~Eun is with Department of Information \& Communication Engineering, DGIST, Korea.}
}


\maketitle

\begin{abstract}
This paper presents a detection algorithm for sensor attacks and a resilient state estimation scheme for a class of uniformly observable nonlinear systems.
An adversary is supposed to corrupt a subset of sensors with the possibly unbounded signals, while the system has sensor redundancy.
We design an individual high-gain observer for each measurement output so that only the observable portion of the system state is obtained.
Then, a nonlinear error correcting problem is solved by collecting all the information from those partial observers and exploiting redundancy.
A computationally efficient, on-line monitoring scheme is presented for attack detection.
Based on the attack detection scheme, an algorithm for resilient state estimation is provided.
The simulation results demonstrate the effectiveness of the proposed algorithm.
\end{abstract}


%
\IEEEpeerreviewmaketitle

\section{Introduction}\label{sec:intro}

Recent developments in network communication and the increase in computational power have made control systems more connected.
As this connectivity increases, the resulting large-scale networked control systems, often referred to as cyber-physical systems, are inherently exposed to the risk of malicious attacks \cite{Sandberg15,Teixeira15,Amin09}.
A variety of attack strategies are adopted by adversaries, and in particular, the sensors themselves or the measurement data of communication networks are often compromised.
For example, the StuxNet worm on the SCADA system \cite{Langner11} and false data injection into power grids \cite{Liu11} have been reported in literature.

To cope with the threats of these attacks,
system designers or defenders have devised sophisticated control algorithms that are more reliable even when some (not all) actuators and measurements are corrupt. 
For example, considering the attack signal as an unknown input, Pasqualetti {\it et al.} \cite{Pasqualetti13} characterized fundamental limitations in attack detection and identification for descriptor linear systems.

Considering only sensor attacks,
the attack identification problem leads to an attack-resilient state estimation problem.
The challenges of this problem are due to the computational complexity being NP-hard \cite{Pasqualetti13} because the general attack identification problem is combinatorial in nature, and hence, solutions require significant computational effort.
Inspired by recent work in the field of compressed sensing \cite{Tao05}, \cite{Donoho06}, Fawzi {\it et al.} \cite{Fawzi14} converted a computationally heavy $\ell_0$-minimization to a convex optimization problem with additional assumptions in their design of an attack-resilient estimator.
An observer-based approach was adopted in \cite{Hespa15ACC}; however, many observers are required to prepare all possible combinations.
A simpler estimation algorithm with a considerably smaller number of observers was independently proposed in \cite{Chanhwa15ECC}, where an observer from each measurement output is constructed and ``partial'' information of the full state is generated.
With all the information collected from each partial observer, the original state can be recovered using an error correction technique.
On the other hand, if a full state observer can be constructed from each sensor, then the method detailed in \cite{Jeon16} can be used.
Their idea is to select correct estimates using a simple median operation, thereby further reducing the computation. 

Although most control systems are nonlinear in practice, all the aforementioned studies are restricted to linear dynamical systems.
An attempt to tackle the resilient state estimation problem for nonlinear systems was first made in \cite{Shou15CDC}, which is a direct extension of the results \cite{Shou15ACC} of linear systems to a class of nonlinear systems, called differentially flat systems.
However, assuming the measurement output to be a ``flat'' output limits the class of systems;
for example, the given system should not have non-trivial zero dynamics \cite{flat15}.
On the other hand, a secure state estimator was constructed in \cite{Hu17} for a special form of nonlinear systems whose stacked outputs can be represented by a linear function of the initial state and the attack vector.

In this paper, we present a dynamic observer-based resilient state estimation scheme under a substantially less restrictive class of uniformly observable nonlinear systems.
Assuming that there are enough number of sensors, we present how to counteract the limited number of sensor attacks.
In particular, $q$-redundant observability (to be defined) is used for attack detection and $2q$-redundant observability for resilient state estimation under $q$-sparse sensor attacks.
The idea of implementation is to design partial observers for each output, as in \cite{Chanhwa15ECC}, which are used to estimate the observable sub-state only and then to process the partial information collected from each sensor.
For this, the ``uniformly observable decomposition'' from \cite{shim2014hybrid}, which is an analogous concept of Kalman observability decomposition for linear systems, and a high gain observer \cite{Gauthier92} are utilized to construct the final detector/estimator.
As a byproduct of high gain observer construction, we obtain an assignable convergence rate of the estimation error that converges to zero.

The proposed attack detection scheme generates a type of residual that is compared with a time-varying threshold.
The required condition for this attack detection is less strong than that for resilient state estimation; this is expected because the attack can also be revealed/identified once the state has been estimated correctly.
It is shown that a detection alarm rings whenever {\em influential} sensor attacks are injected.
By ``influential,'' we mean that if the alarm does not ring, then either there is no attack or the attack is so small that it cannot be distinguished from measurement noise/disturbance.
Finally, by employing the time-varying threshold, the proposed scheme also considers the transient of the estimation error caused by the dynamic observers.
The proposed attack detection algorithm enables resilient state estimation by signaling corruption in the current combination of sensor information.
In this way, one can avoid solving an optimization problem at each sampling time as in \cite{Fawzi14,Hespa15ACC,Chanhwa15ECC}.
The preliminary version of this paper has been presented in \cite{Kim16CDC}.

\section{Problem Formulation and Preliminaries}\label{sec:preliminary}

\subsection{Problem Formulation}

We consider a smooth nonlinear system given by
\begin{subequations}\label{eq:system}
	\begin{equation}
	\dot x(t) = f(x(t))+g(x(t))u(t) \label{eq:systemDynamics}
	\end{equation}
where $x \in \R^n$ is the state and $u \in \R$ is the input.
We assume that the state and the input of system \eqref{eq:system} are bounded.
More specifically, $u(t) \in \UU$ for all $t \ge 0$ where $\UU$ is a compact set, and $x(t) \in \XX := \{ w \in \R^n : \|w\|_\infty \le M_x \}$, $t \ge 0$, with a constant $M_x > 0$, where $\|w\|_\infty := \max_{1\le i \le n} |w_i|$.

Suppose that there are $p$ sensors to measure (a smooth function of) the state, which is vulnerable to sensor attack:
	\begin{align}\label{eq:systemOutputIndividual}
	y_i(t) = h_i(x(t)) + a_i(t)+v_i(t), ~ i \in [p] := \{1,\cdots,p\}
	\end{align}
\end{subequations}
where $y_i \in \R$ is the value of sensor $i$, $a_i \in \R$ is the attack signal injected to the sensor, and $v_i\in \R$ is the measurement noise.
Throughout the paper, the measurement noise $v_i(t)$ is assumed to be bounded by a constant $M_{v,i}$ for each $i\in[p]$.
On the other hand, the attack signal $a_i(t)$ is not assumed to be a bounded signal, and a craftily designed $a_i(t)$ can corrupt $y_i(t)$ to have an arbitrary value.
This fact makes it difficult to detect attacks, and more difficult to estimate the state $x$ from the measured outputs. 

Instead of imposing restrictions on the attack signal $a_i(t)$ itself, we assume $q$-sparsity on the set of attack signals $\{a_i\}_{i\in[p]}$, motivated by the rationale that the attack resource is limited so that only a portion of the sensors is compromised (see \cite{Fawzi14, Hespa15ACC, Chanhwa15ECC, Shou15CDC, Shou15ACC}).

\begin{asm}\label{asm:sparsity}
Up to $q$ ($q < p$) sensors are compromised.
In other words, with the index set $U$ of uncompromised sensors, defined by $U := \{ i \in [p] : a_i \equiv 0 \}$, it is assumed that $|U| \ge p-q$ where $|U|$ is the cardinality of the set $U$.
The set $U$ is unknown (to the defender).
\end{asm}

Based on this assumption, two problems are of interest in this paper.
The first is the real-time {\em detection of the sensor attack}
from only the information of the system model \eqref{eq:system}, input $u$, and the outputs $\{y_i\}_{i\in[p]}$ up to time $t$.
It is shown that this problem is solved if system \eqref{eq:system} satisfies ``$q$-redundant observability,'' which basically implies observability of \eqref{eq:system}, even when any $q$ sensors (out of $p$ sensors) are removed.
The second problem is the generation of a signal $\hat x(t)$ that converges to the true state $x(t)$ despite the attack satisfying Assumption~\ref{asm:sparsity}.
This problem is called {\em the resilient state estimation}, and to solve this, we require a stronger condition of ``$2q$-redundant observability'' for system \eqref{eq:system}.

\subsection{Preliminaries}

Our idea for solving these problems is to construct $p$ nonlinear observers to each individual output $y_i$, $i \in [p]$.
As there is no guarantee that the state $x$ is observable from the single output $y_i$, each observer cannot recover the full state $x$ in general. 
Instead, each observer can recover an observable portion of the state only. 
By observable portion, we mean the observable substate at a special coordinate.
For linear systems, this substate corresponds to the observable subsystem in the well-known Kalman observable decomposition.
For nonlinear systems, we assume that the observable subsystem is ``uniformly observable\footnote{Unlike linear systems, a nonlinear system can be both observable and unobservable depending on the input signal $u(t)$ in general. In contrast, a uniformly observable nonlinear system is observable for every input; i.e., observable uniformly in inputs. See \cite{Gauthier92} for details. This stronger notion allows nonlinear observer design in most cases.} \cite{Gauthier92}''.
The following assumption asks uniform observability of the observable portion of system \eqref{eq:system} for the individual output $y_i$.

\begin{asm}\label{asm:uniformObservableDecomposition}
	For each $i \in [p]$, there exist a natural number $n_i$ and a diffeomorphism $\Xi_i : \R^n \ra \R^{n_i}\times\R^{n-n_i}$ such that, by $[z_i^T,z_i'^T]^T := \Xi_i(x)$, system \eqref{eq:system} is transformed into the form
	\begin{subequations}\label{eq:transformed}
		\begin{align}
		\dot z_i &= \begin{bmatrix} \dot z_{i,1} \\ \dot z_{i,2} \\ \vdots \\ \dot z_{i,n_i} \end{bmatrix}
		= \begin{bmatrix} z_{i,2} \\ \vdots \\ z_{i,n_i} \\ \alpha_{i}(z_i) \end{bmatrix}
		+ \begin{bmatrix} \beta_{i,1}(z_{i,1}) \\ \beta_{i,2}(z_{i,1}, z_{i,2}) \\ \vdots \\\beta_{i,n_i}(z_{i,1}, \cdots, z_{i,n_i}) \end{bmatrix}u \label{eq:transformedObservablePart} \\
		\dot z_i' &= F_i'(z_i,z_i') + G_i'(z_i,z_i')u \\
		y_i &= z_{i,1} + a_i + v_i . \label{eq:transformedOutput}
		\end{align}
	\end{subequations}
	Moreover, the functions $\alpha: \R^{n_i} \to \R$ and $\beta_{i,j}: \R^{j} \to \R$, $j \in [n_i]$, are globally Lipschitz.
\end{asm}

A few comments follow regarding Assumption \ref{asm:uniformObservableDecomposition}.
\begin{enumerate}
	\item The first $n_i$ component of the diffeomorphism $\Xi_i$ is given by $[h_i, L_fh_i, L_f^2h_i, \cdots, L_f^{n_i-1}h_i]^T$; i.e.,
	\begin{align}\label{eq:observablePartMapping}
	z_i = \begin{bmatrix} z_{i,1} \\ z_{i,2} \\ \vdots \\ z_{i,n_i} \end{bmatrix} = 
	\begin{bmatrix}
	h_i(x) \\ L_f h_i(x) \\ \vdots \\ L_{f}^{n_i -1}h_i(x)
	\end{bmatrix} =: \Phi_i(x)
	\end{align}
	which is easily verified by comparing \eqref{eq:system} and \eqref{eq:transformed} with $u \equiv 0$.
	
	\item The substate $z_i$ corresponds to the observable substate from the output $y_i$, whereas $z_i'$ is the unobservable substate.
	This is obvious from the structure of \eqref{eq:transformed}.
	On the other hand, the triangular structure of $\beta_i = [\beta_{i,1}, \cdots, \beta_{i,n_i}]^T$ is a necessary and sufficient condition for uniform observability of the $z_i$-subsystem (see \cite{Gauthier92} for the proof of this statement).
	
	\item\label{item:extension} Requesting global Lipschitz properties for $\alpha_i$ and $\beta_{i.j}$, $j \in [n_i]$, is not a restriction owing to the boundedness of $x(t)$.
	Indeed, noting that $x(t) \in \XX$, one can find a constant $M_{z,i}$ such that $\|z_i\|_\infty = \| \Phi_i(x) \|_\infty \le M_{z,i}$ for all $x \in \XX$.
	Then, one can modify $\alpha_i$ and $\beta_{i,j}$ outside the set ${\mathcal Z}^i := \{z_i : \|z_i\|_\infty \le M_{z,i} \}$ so that $\alpha_i$ and $\beta_{i,j}$ are globally Lipschitz while they remain the same in ${\mathcal Z}^i$.
	In theory, this modification is always possible by Kirszbraun's Lipschitz extension theorem\footnote{For a function $f:X \to \R$ that is Lipschitz on $X$, a Lipschitz extension is given by $\overline{ f}(x) := \inf_{y \in X}( f(y) + \overline{\mathrm{Lip}}(f) |x - y|)$ where $\overline{\mathrm{Lip}}(f)$ is a Lipschitz constant of $f$ on $X$. 
	For a vector-valued function $f$, this extension is applied to each component.
	} \cite[p.~21]{Schwartz69}.
	In practice, a simpler method can be used. 
	For example, $\alpha_i(z_i)$ is replaced by
	\begin{equation}\label{eq:LipschitzExtensionSimpler}
	\overline{ \alpha_i}(z_i) = \alpha_i(\sat(z_i,M_{z,i}))
	\end{equation}
	where $\sat$ is the component-wise saturation function; i.e.,
	for $w =[w_1,\cdots,w_k]^T\in \R^k$ and $M>0$,  
	$$\sat(w,M) := \begin{bmatrix}
	\min\{\max\{w_1,-M\},M\}\\\vdots\\\min\{\max\{w_k,-M\},M\}
	\end{bmatrix} \in \R^k.$$
	See \cite[Sec.~3.3]{shim2000passivity} for more details.
	
	\item For linear systems, Assumption \ref{asm:uniformObservableDecomposition} always holds.
\end{enumerate}

\medskip

For each $i \in [p]$, a partial high gain observer only for observable part \eqref{eq:transformedObservablePart} and \eqref{eq:transformedOutput} is constructed by
\begin{align}
\dot{\hat{z}}_{i}
= \begin{bmatrix}\dot{\hat{z}}_{i,1}\\ \dot{\hat{z}}_{i,2}\\ \vdots\\ \dot{\hat{z}}_{i,n_i}\end{bmatrix}
&= \begin{bmatrix} {\hat{z}_{i,2}}\\ \vdots \\{\hat{z}_{i, n_i}}\\ \alpha_{i}({\hat{z}_{i}})\end{bmatrix}
+ \begin{bmatrix} \beta_{i,1}({\hat{z}_{i,1}})\\ \beta_{i,2}({\hat{z}_{i,1}}, {\hat{z}_{i,2}}) \\ \vdots \\\beta_{i,n_i}({\hat{z}_{i,1}}, \cdots, {\hat{z}_{i, n_i}})\end{bmatrix}u\notag\\
&\qquad -P_{i}^{-1}C_{i}^{T} (C_{i} \hat{z}_i-y_i)\label{eq:gauthierObserver}
\end{align}
where $\hat z_i$ is the estimated state of $z_i$, $C_{i}:= [1,0,\cdots,0]\in \R^{1\times n_i}$, and $P_i(\theta_i) \in \R^{n_i\times n_i}$ is the unique positive definite solution of
$$0 = -\theta_i P_{i} - A_{i}^T P_{i} - P_{i}A_{i} + C_{i}^T C_{i}$$
where $\theta_i$ is a constant to be determined, and $A_i \in \R^{n_i \times n_i}$ is a matrix whose $(i,j)$-th element is $1$ if $i+1 = j$ and $0$ otherwise.
We suppose that the initial condition is set such that $\| \hat z_i(0) \|_\infty \le M_{z,i}$.
The parameter $\theta_i$ is determined by the following lemma (while $\theta_i$ is often taken as a sufficiently large number obtained from repeated simulations in practice).

\begin{lem}\label{lem:gauthierObserver}{\bf (\!\!\cite{Gauthier92}, \cite{shim2000passivity}) }
There exist positive constant $\theta_i^* \ge 1$ and non-decreasing functions $\eta_i(\theta_i)$ and $\epsilon_i(\theta_i)$ such that, for each $\theta_i \ge \theta_i^*$, the observer \eqref{eq:gauthierObserver} guarantees
\begin{equation}\label{eq:partialObsConvergence}
\hspace{-2mm}	\| \hat z_i(t) - z_i(t) \|_\infty \le \max\{ \eta_i(\theta_i) e^{- \frac{\theta_i}{8}t} \|\hat z_i(0)- z_i(0) \|_\infty,\epsilon_i(\theta_i)\}, 
\end{equation}
provided $a_i(t)=0$ for all $t\ge 0$.
\end{lem}

{\em Proof:}
The proof is given in Appendix.
\hfill$\blacksquare$

The estimate $\hat z_i$ from a compromised observer (i.e., $a_i(t)\not\equiv 0$) is not expected to satisfy \eqref{eq:partialObsConvergence}, and may have arbitrarily large values when the attack signal is unbounded.
In order to prevent $\hat z_i(t)$ from diverging to infinity during the attack on sensor $i$, let us introduce the following reset rule\footnote{This reset rule does not yield ``zeno behavior,'' i.e., it does not make infinitely many resets in finite time, unless the attack signal $a_i(t)$ tends to infinity in finite time.} for the observer \eqref{eq:gauthierObserver}:
\begin{equation}\label{eq:resetRule}
\hat z_i(t^+) \leftarrow \hat z_0 \quad \text{if}~ \|\hat z_i(t)\|_\infty > \max\{ 2\eta_i M_{z,i},\epsilon_i\}+M_{z,i},
\end{equation}
where $\hat z_0$ is any vector such that $\|\hat z_0\|_\infty \le M_{z,i}$.
This rule is inspired by the fact that, if $a_i(t) \equiv 0$, then Lemma \ref{lem:gauthierObserver} guarantees that $\| \hat z_i(t)\|_\infty \le \| \hat z_i(t) - z_i(t) \|_\infty + \| z_i(t) \|_\infty \le\max\{ 2\eta_i M_{z,i}, \epsilon_i\} + M_{z,i}$.
Therefore, the proposed observer \eqref{eq:gauthierObserver} with \eqref{eq:resetRule} guarantees that $\| \hat z_i(t) \|_\infty \le \max\{ 2\eta_i M_{z,i},\epsilon_i\}+M_{z,i}$ for all $i \in [p]$ and for all $t$, which is used in Section III.D to relax Assumption \ref{asm:sparsity}.

\section{Main Results}

Lemma \ref{lem:gauthierObserver} ensures $\limsup_{t \to \infty} \|\hat z_i(t) - z_i(t)\|_\infty \le \epsilon_i$ provided there is no attack on sensor $i$.
However, if the output $y_i$ of sensor $i$ is compromised by an attack, the estimate $\hat z_i(t)$ behaves unpredictably and useful information cannot be obtained from $\hat z_i(t)$.
Fortunately, under Assumption \ref{asm:sparsity}, at least $p-q$ estimates in $\{ \hat z_i(t) \}_{i \in [p]}$ remain uncompromised.
It is noted that the benefit of installing partial observers to individual outputs $y_i$, rather than a single full observer for the full set $\{y_i\}_{i\in[p]} $ of measurements, is clear; the effect of attack signal $a_i$ is restricted to $\hat z_i$ and does not propagate to other estimates.
Thus, our task is to determine the uncompromised estimates and recover the state $x(t)$ from them.
In the forthcoming discussion, we deal with partitions of vectors frequently; thus, let us define some notation and terminology that facilitate this discussion.

\subsection{Notation and Terminology}\label{sec:notation}

Recall that $[p]$ denotes the set of natural numbers from $1$ to $p$ as $[p]=\{1,2,\cdots,p\}$.
For a finite sequence $\NN$ of $p$ natural number $n_i$'s, i.e., ${\NN} = (n_1,n_2,\cdots,n_p)$, $\R^{\NN}$ is defined as the Cartesian product of Euclidean spaces:
\begin{align*}
\R^{\NN} &= \R^{(n_1,n_2,\cdots,n_p)} = \R^{n_1} \times \R^{n_2} \times \cdots \times \R^{n_p} \\
&= \{ (z_{1}, z_{2}, \cdots, z_{p}) : z_{i} \in \R^{n_{i}},\,i\in [p] \}.
\end{align*}
While this space can be identified as a single Euclidean space of dimension $N = \sum_{i=1}^p n_i$; i.e.,
\begin{align*}
\R^{\NN} = \R^{N},\quad (z_{1}, z_{2}, \cdots, z_{p}) = [z_1^T, z_{2}^T, \cdots, z_{p}^T]^T,
\end{align*}
we consider $\R^{\NN}$ more often than $\R^N$.
A subsequence of $\NN$ with indices in a subset $I = \{ i_1< i_2< \cdots< i_l : i_j \in [p] \}$ is denoted by $\NN_I := (n_{i_1}, n_{i_2}, \cdots, n_{i_l})$.
With the index set $I$, a canonical projection
\begin{align}
\begin{split}\label{eq:notation1}
\pi_I : \R^{\NN} &\ra \R^{\NN_I}\\
(z_{1}, z_{2}, \cdots, z_{p}) &\mapsto (z_{i_1}, z_{i_2}, \cdots, z_{i_l})
\end{split}
\end{align}
selects $l$ elements out of $p$ tuples.
For a given vector $z\in\R^{\NN}$ and for a given function $\Phi : X \ra \R^{\NN}$, respectively, we define
\begin{align*}
z_I &:= \pi_I(z) & &\text{so that $z_I \in \R^{\NN_I}$}, \\
\Phi_I &:= \pi_I \circ\Phi & &\text{so that $\Phi_I : X \ra \R^{\NN_I}$}.
\end{align*}
Let $|I|$ for an index set $I$ be the cardinality of $I$, and define a collection of index subsets as $\binom{I}{q} := \{J\subset I : \left |J \right| = q \}$ with a nonnegative integer $q \le \left| I \right |$.
Finally, for a given sequence of natural numbers $\NN$ of length $p$ and for $w = (w_1,w_2,\cdots,w_p)$ $\in \R^{\NN}$, define $\|w\|_0^{\NN} = | \{ i\in[p] : w_{i} \neq 0 \} |$ (by abusing the conventional $l_0$-norm notation), and the vector $w$ is said to be {\em $q$-sparse} if $\|w\|_0^\NN \le q$; that is, there exists an index set $I \in \binom{[p]}{p-q} := \{J\subset [p] : \left |J \right| = p-q \} $ such that $w_I = 0$.

The notion of bi-Lipschitz function and its left inverse is actively used in this paper.
With $X \subset \R^n$, a mapping $\phi:X\ra \R^m$ is {\em Lipschitz} (on $X$) if there exists a constant $\overline{L}$ such that 
$$\| \phi(x_1) - \phi(x_2)\|_\infty \le \overline{L}\|x_1 - x_2 \|_\infty, \quad \forall x_1, x_2 \in X.$$
The infimum of such $\overline{L}$ is indicated as $\overline{\mathrm{Lip}}(\phi)$.
It is {\em bi-Lipschitz} (on $X$) if, in addition, there exists a positive constant $\underline{L}$ such that
$$\underline{L}\|x_1 - x_2 \|_\infty \le \| \phi(x_1) - \phi(x_2)\|_\infty, \quad \forall x_1, x_2 \in X.$$
The supremum of such $\underline{L}$ is indicated as $\underline{\mathrm{Lip}}(\phi)$.
For a given bi-Lipschitz function $\phi: X \ra \R^m$, a function $\psi: \R^m \ra X$ is called a {\em Lipschitz-extended left inverse of $\phi$} if it is defined and Lipschitz {\em on $\R^m$} and satisfies $\psi(\phi(x)) = x$ for all $x \in X$. 
It is obvious that a bi-Lipschitz map is injective, and thus its inverse exists on its image $Y=\phi(X)$ and the inverse is Lipschitz on $Y$.
However, it should be noted that the Lipschitz-extended left inverse $\psi$ is defined on the whole codomain $\R^m$ and its image $\psi(\R^m)$ is $X \subset \R^n$.
The identity function on the set $X$ is denoted by $\id_{X}$.

A differentiable function $\phi: X\ra \R^m$ is called an {\it immersion} if its Jacobian matrix has full column rank for every $x \in X$.

\begin{prop}\label{prop:locallyBi-Lipschitz}
If $\phi$ is an injective immersion on a compact set $X$, then it is bi-Lipschitz on $X$.
\end{prop}

{\em Proof:}
The proof is given in Appendix.
\hfill$\blacksquare$

\smallskip

For example, if $X=[-1,1] \times [-1,1] \subset \R^2$ and $\phi(x) = Tx$ with a matrix $T \in \R^{3 \times 2}$ of full column rank, then $\phi$ is an injective immersion and thus it is a bi-Lipschitz function on $X$.
One of its Lipschitz-extended left inverses is given by $\psi(z) = \sat(T^\dagger z,1)$ where $T^\dagger \in \R^{2 \times 3}$ is a left inverse matrix of $T$, which maps $\R^3$ to $X$.

\subsection{Redundant Observability}

Let us rewrite the equations in \eqref{eq:observablePartMapping} for all $i \in [p]$ simultaneously as
\begin{equation}\label{eq:Phi}
z := \begin{bmatrix} z_1 \\ \vdots \\ z_p \end{bmatrix} = \begin{bmatrix} \Phi_1(x) \\ \vdots \\ \Phi_p(x) \end{bmatrix} =: \Phi(x)
\end{equation}
so that the stack $z$ is defined as a partitioned vector in $\R^\NN$ where $\NN = (n_1,n_2,\cdots,n_p)$.
To recover the state $x$ from the collection of estimates $\hat z := (\hat z_1, \hat z_2,\cdots,\hat z_p)$ in \eqref{eq:gauthierObserver}, the function $\Phi: \XX \to \Phi(\XX) \subset \R^{\NN}$ should have injectivity so that it has a left inverse $\Phi^{-1}$, defined at least on its image $\Phi(\XX)$. Let the estimate of $x$ be the left inverse of $\hat z$ if $\hat z = z$.
In addition to injectivity, we require that the mapping $\Phi$ is an immersion in order to ensure bi-Lipschitzness (which is used later) on the domain $\XX$.
Asking $\Phi$ to be an injective immersion is, in fact, an extension of the linear case since the Jacobian of $\Phi$ corresponds to the observability matrix,
which has full column rank.
Moreover, since up to $q$ estimates in $\{ \hat z_i \}_{i \in [p]}$ might be compromised, we require some redundancy in the map $\Phi$ to ensure that the map remains an injective immersion if any $q$ components $\Phi_i$ are eliminated from $\Phi$.
The following definition states this requirement precisely.

\begin{defn}\label{def:redundantInjectiveImmersion}
	System \eqref{eq:system} is said to be {\em $k$-redundant observable} if, for the mapping $\Phi: \XX \to \R^{\NN}$ in \eqref{eq:Phi}, 
	the function $\Phi_I: \XX \to \R^{\NN_I}$ is an injective immersion for all $I \subset [p]$ such that $| I | = p-k$.
\end{defn}

It is straightforward to see that $k$-redundant observability implies $k'$-redundant observability for any $k' < k$, and 
$0$-redundant observability can be regarded as conventional observability of \eqref{eq:system}.

Now, it is noted that, although $k$-redundant observability of \eqref{eq:system} guarantees the existence of a left inverse $\Phi_I^{-1}$ of $\Phi_I$ where $I \in \binom{[p]}{p-k}$, the inverse $\Phi_I^{-1}$ is defined only on $\Phi_I(\XX)$.
While it is true that $z_I(t) \in \Phi_I(\XX) \subset \R^{\NN_I}$, there is no guarantee that the estimate $\hat z_I(t)$, that converges to $z_I(t)$, belongs to $\Phi_I(\XX)$.
To use the left inverse of $\Phi_I$ on the whole space $\R^{\NN_I}$, let us define our {\em Lipschitz-extended left inverse} of $\Phi_I$ as
\begin{align}\label{eq:leftinverse}
\begin{split}
\Psi^I : \R^{\NN_I} &\to \XX\\
\zz_I ~~&\mapsto \sat(\overline{\Phi_I^{-1}}(\zz_I) ,~M_x) 
\end{split}
\end{align}
in which, $\overline{\Phi_I^{-1}}$ is a Lipschitz extension of $\Phi_I^{-1}$ from $\Phi_I(\XX)$ to $\R^{\NN_I}$ (refer to Item \ref{item:extension} following Assumption \ref{asm:uniformObservableDecomposition}), and the saturation function is employed to map the image of $\overline{\Phi_I^{-1}}$ into the set $\XX$.
Indeed, this function $\Psi^I$ is globally Lipschitz on $\R^{\NN_I}$ because $\Phi_I$ is bi-Lipschitz on $\XX$ by Proposition \ref{prop:locallyBi-Lipschitz}; and thus, a left inverse of $\Phi_I$ exists on $\Phi_I(\XX)$ which is Lipschitz on $\Phi_I(\XX)$. It is then extended to be globally Lipschitz on $\R^{\NN_I}$, and the saturation function at the end preserves the Lipschitz property.
With the global Lipschitz inverse function $\Psi^I$ at hand, let an estimate of $x(t)$ be
$$\hat x^I(t) := \Psi^I(\hat z_I(t)) \quad \in \XX.$$
\begin{rem}\label{rem:LipschitzLeftInverseConstruction}
	For the simple construction of the Lipschitz extension $\overline{\Phi_I^{-1}}$ in practice,
	one may want to employ a method using saturation functions as in \eqref{eq:LipschitzExtensionSimpler}.
	Let $M_z:=\max_{i\in [p]}M_{z,i}$, and $\mathcal{Z}^I:= \{z_I\in\R^{\NN_I}: \|z_I\|_\infty\le M_z\}$ which contains $\Phi(\XX)$ by construction.
	If there is a smooth function $\Phi_I'^{-1}$ defined on $\mathcal{Z}^I$ such that $\Phi_I'^{-1}(\zz_I) = \Phi_I^{-1}(\zz_I)$ for all $\zz_I \in \Phi(\XX)$, then
	a Lipschitz extension $\overline{\Phi_I^{-1}}$ is easily obtained by
	\begin{equation}\label{eq:leftinverseReplaced}
	\overline{\Phi_I^{-1}}(\zz_I)=\Phi_I'^{-1}(\sat(\zz_I,\, M_z))).
	\end{equation}
\end{rem}

Suppose that system \eqref{eq:system} is $q$-redundant observable.
Since up to $q$ sensors are compromised, there is at least one index set $I \subset [p]$ with $|I|=p-q$ such that $I$ is contained in the set $U$ of Assumption \ref{asm:sparsity}.
In this case, by Lemma~\ref{lem:gauthierObserver}, we have
\begin{align*}
\| \hat x^I(t) - x(t) \|_\infty &= \| \Psi^I(\hat z_I(t)) - \Psi^I(z_I(t)) \|_\infty \\
&\le \overline{\mathrm {Lip}}(\Psi^I) \max\{\gamma(t),\epsilon\}
\end{align*}
in which
\begin{equation}\label{eq:epsilon_delta}
\gamma(t):= \max_{i\in[p]}\{2 M_{z,i} \eta_i(\theta_i) e^{- \frac{\theta_i}{8}t}\},\quad \epsilon:= \max_{i\in[p]}\{\epsilon_i(\theta_i)\}.
\end{equation}
Thus, $x(t)$ is recovered by $\hat x^I(t)$ and the error is eventually bounded by $\overline{\mathrm{Lip}}(\Psi^I)\epsilon$, which is an upper error bound caused by the measurement noise.
The remaining question is that, since the set $U$ is unknown, how to find $I$ such that $I \subset U$.

\subsection{Detection of Sensor Attack}
We begin by observing that the difference between $\hat z(t)$ and $z(t) = \Phi(x(t))$ is written as
\begin{equation}\label{eq:staticEq}
\hat z(t) - \Phi(x(t)) = e(t) + r(t) \quad \in \R^{\NN}
\end{equation}
where the vector $e$ represents the error caused by the injected sensor attack, and the vector $r$ is the estimation error in the partial observers.
Thus, if there is no attack, then $e(t) \equiv 0$, and the norm of $r_i(t) = \hat z_i(t) - \Phi_i(x(t)) \in \R^{n_i}$ decreases as in \eqref{eq:partialObsConvergence} of Lemma \ref{lem:gauthierObserver} for all $i \in [p]$.
Under Assumption \ref{asm:sparsity}, $e_U(t) \equiv 0$, and thus, $\|r_U(t)\|_\infty = \|\hat z_U(t) - \Phi_U(x(t))\|_\infty$ is eventually bounded by the constant $\epsilon$ from \eqref{eq:epsilon_delta}.
In contrast, the vector $e_{[p]-U}(t)$ may not be zero and the estimation error $\hat z_{[p]-U}(t) - z_{[p]-U}(t)$ may become large.
Since we have no restriction on the value of $e_{[p]-U}$, it is equivalent to treat $e_{[p]-U}$ and $r_{[p]-U}$ as
$$e_{[p]-U} = \hat z_{[p]-U} - \Phi_{[p]-U}(x) \quad \text{and} \quad r_{[p]-U} = 0$$
because it holds that $\hat z_{[p]-U}(t) - \Phi_{[p]-U}(x(t)) = e_{[p]-U}(t) + r_{[p]-U}(t)$. Moreover, we have
\begin{equation}\label{eq:errorProperties}
\|e(t)\|^{\NN}_0 \le q,\quad
\|r(t)\|_\infty \le \delta(t):= \max\{\gamma(t),\epsilon\}
\end{equation}
from Lemma \ref{lem:gauthierObserver} and Assumption \ref{asm:sparsity}.

Now, the following theorem presents a detection mechanism for influential attacks.

\begin{thm}\label{thm:staticDetection}
Under Assumptions \ref{asm:sparsity} and \ref{asm:uniformObservableDecomposition}, assume that system \eqref{eq:system} is $2q$-redundant observable.
For a given $I \subset [p]$ with $|I| = p-q$, consider an inequality for the distance from $\hat z_I$ to a point in the image of $\Phi_I$:
\begin{equation}\label{eq:errordetection}
\| \hat z_I(t) - \Phi_I(\Psi^I(\hat z_I(t))) \|_\infty > \overline{\mathrm{Lip}}(\id_{\R^{\NN_I}} - \Phi_I\circ\Psi^I) \delta(t).
\end{equation}
\begin{enumerate}
	\item\label{thm:1.1} If \eqref{eq:errordetection} holds, then $e_I(t) \not = 0$; i.e., there is a sensor attack among the sensors whose indices $i$ belong to $I$.
	\item\label{thm:1.2} If \eqref{eq:errordetection} is violated, then
	\begin{equation*}	
	\| \hat x^I(t) - x(t) \|_\infty \le \frac{\overline{\mathrm{Lip}}(\id_{\R^{\NN_I}} - \Phi_I\circ\Psi^I)+1 }{\min_{\{J \subset I: |J| = p-2q \}} \underline{\mathrm{Lip}}(\Phi_J)} \delta(t) .
	\end{equation*}
\end{enumerate}
\end{thm}

{\em Proof:}
\ref{thm:1.1}) It follows that 
\begin{align}
&\| \hat z_I - \Phi_I(\Psi^I(\hat z_I)) \|_\infty\notag\\
&= \| (\hat z_I - \Phi_I(\Psi^I(\hat z_I)))-( \Phi_I(x) - \Phi_I(\Psi^I( \Phi_I(x)))) \|_\infty \notag\\
&=\|(\id_{\R^{\NN_I}} - \Phi_I\circ\Psi^I)(\hat z_I) - (\id_{\R^{\NN_I}} - \Phi_I\circ\Psi^I)(\Phi_I(x))\|_\infty \notag\\
&\le \overline{\mathrm{Lip}}(\id_{\R^{\NN_I}} - \Phi_I\circ\Psi^I)\| \hat z_I - \Phi_I(x)\|_\infty\label{eq:ThmPf1}\\
&= \overline{\mathrm{Lip}}(\id_{\R^{\NN_I}} - \Phi_I\circ\Psi^I)\| e_I + r_I \|_\infty.\notag
\end{align}
Hence, if $e_I(t) = 0$, then
$$\| \hat z_I(t) - \Phi_I(\Psi^I(\hat z_I(t))) \|_\infty \le \overline{\mathrm{Lip}}(\id_{\R^{\NN_I}} - \Phi_I\circ\Psi^I) \delta(t)$$
as $\|r_I(t)\|_\infty \le \delta(t)$.
This proves the claim.

\ref{thm:1.2}) As $e_I$ is $q$-sparse (by the fact that $e$ is $q$-sparse), i.e., $\|e_I\|_0^{\NN} \le q$, in which $|I|=p-q$, there is an index set $J \subset I$ such that $|J| = p-2q$ and $e_J = 0$.
Then, it follows from \eqref{eq:staticEq} that $\hat z_J = \Phi_J(x) + r_J$ and we have 
\begin{align*}
\|\hat{z}_I - \Phi_I(\Psi^I(\hat{z}_I)) \|_\infty &\ge \|\hat{z}_J - \Phi_J(\Psi^I(\hat{z}_I)) \|_\infty\\
&= \|\Phi_J(x) + r_J - \Phi_J(\hat x^I) \|_\infty \\
&\ge {\underline{\mathrm{Lip}}(\Phi_J)} \|x - \hat x^I \|_\infty - \|r_J \|_\infty\\
&\ge {\underline{\mathrm{Lip}}(\Phi_J)} \|x - \hat x^I \|_\infty - \delta.
\end{align*}
Therefore, when \eqref{eq:errordetection} is violated, we obtain
$${\underline{\mathrm{Lip}}(\Phi_J)} \|\hat x^I - x \|_\infty - \delta \le \overline{\mathrm{Lip}}(\id_{\R^{\NN_I}} - \Phi_I\circ\Psi^I)\delta.$$
From $2q$-redundant observability, it follows that ${\underline{\mathrm{Lip}}(\Phi_J)} > 0$ for any $J$ such that $|J|=p-2q$ as $\Phi_J$ is bi-Lipschitz on $\XX$ according to Proposition \ref{prop:locallyBi-Lipschitz}.
This completes the proof.
\hfill$\blacksquare$

\begin{cor}\label{cor:1}
Under Assumptions \ref{asm:sparsity} and \ref{asm:uniformObservableDecomposition}, assume that system \eqref{eq:system} is $q$-redundant observable.
Consider
\begin{equation}\label{eq:errordetection2}
\| \hat z(t) - \Phi(\Psi(\hat z(t))) \|_\infty > \overline{\mathrm{Lip}}(\id_{\R^{\NN}} - \Phi\circ\Psi) \delta(t)
\end{equation}
where $\Psi$ is a Lipschitz-extended left inverse of $\Phi$.
\begin{enumerate}
	\item If \eqref{eq:errordetection2} holds, then $e(t) \not = 0$; i.e., there is a sensor attack.
	\item If \eqref{eq:errordetection2} is violated, then
	\begin{equation*}		
	\| \hat x(t) - x(t) \|_\infty \le \frac{\overline{\mathrm{Lip}}(\id_{\R^{\NN}} - \Phi\circ\Psi)+1 }{\min_{\{J \subset [p]: |J| = p-q \}} \underline{\mathrm{Lip}}(\Phi_J)} \delta(t)
	\end{equation*}
	where $\hat x(t) := \Psi(\hat z (t))$.
\end{enumerate}	
\end{cor}

{\em Proof:} 
This follows from the proof of Theorem \ref{thm:staticDetection} in which $I$ is replaced by $[p]$, and the condition $|J|=p-2q$ is replaced by the condition $|J|=p-q$. 
Hence, $q$-redundant observability is sufficient in this case.
\hfill$\blacksquare$

\medskip

Inequality \eqref{eq:errordetection2} is the key to sensor attack detection.
It is noted that both sides of \eqref{eq:errordetection2} can be readily evaluated as all the quantities are available at all time $t \ge 0$.
By checking \eqref{eq:errordetection2}, one can detect a sensor attack.
Of course, violation of \eqref{eq:errordetection2} does not necessarily imply no sensor attack.
However, even when there is an attack, its effect on the state estimation is limited as seen in the theorem, as $\delta$ is usually small.
This case occurs when the size of error $e$ is so small that distinction between the error $r$ caused by measurement noise and the error $e$ caused by the attack is not possible.

Note that Theorem \ref{thm:staticDetection} explains the sensor attack detection for a given subset $I$,
whereas Corollary \ref{cor:1} detects sensor attacks for the whole set $[p]$. 
Thus, the discussion in \eqref{eq:errordetection2} also applies to \eqref{eq:errordetection}.
When \eqref{eq:errordetection} is violated, we suppose that there is no influential attack in $\{y_i\}_{i\in I}$, and the signals $\{ \hat z_i \}_{i \in I}$ are trustworthy.
By repeating \eqref{eq:errordetection} with all subsets $I \subset [p]$ satisfying $|I| = p-q$, one can always find a trustworthy set of sensors as at most $q$ sensors are compromised.
This is the main idea of the resilient state estimation scheme presented in the next subsection, and the following remark justifies why $2q$-redundant observability is required in Theorem \ref{thm:staticDetection} unlike in Corollary \ref{cor:1}.

\begin{rem}
The reason why $2q$-redundant observability, which is stronger than $q$-redundant observability, is needed in Theorem \ref{thm:staticDetection} is as follows.
Suppose that there is $x' \in \XX$ such that $x' \not = x$ and $\Phi_I(x')-\Phi_I(x)$ is $q$-sparse and not necessarily small.
If a $q$-sparse attack $e_I$ happens to be the same as $\Phi_I(x')-\Phi_I(x)$, i.e., $e_I = \Phi_I(x')-\Phi_I(x)$, then, even when the estimation error $r_I$ is zero, we have $\| \hat z_I - \Phi_I(\Psi^I(\hat z_I)) \|_\infty = 0$ because $\hat z_I = \Phi_I(x) + e_I = \Phi_I(x')$.
Hence, condition \eqref{eq:errordetection} cannot detect the attack.
Fortunately, this pathological case does not occur owing to $2q$-redundant observability.
Indeed, with $J \subset I$ such that $|J| = p-2q$ and $e_J=0$, we have $\Phi_J(x') = \Phi_J(x)$.
As $\Phi_J$ is injective, it follows that $x'=x$.
This is the underlying idea of the proof of Theorem \ref{thm:staticDetection}.
\end{rem}

\subsection{Resilient State Estimation}

To present the proposed resilient state estimation scheme in a more practical situation, let us assume that the sensors compromized by the attack can change from time to time; that is, we assume a relaxed version of Assumption \ref{asm:sparsity} as follows.

\begin{asm}\label{asm:sparsity2}
Let $\Delta_1$ and $\Delta_2$ be sufficiently large constants such that
\begin{subequations}\label{eq:Delta}
\begin{align}
(\max\{2\eta_i M_{z,i},\epsilon_i\}+2M_{z,i})\eta_i e^{-\frac{\theta_i}{8}\Delta_1}&\le \epsilon_i \label{eq:Delta1} \\
2M_{z,i}\eta_i e^{-\frac{\theta_i}{8}\Delta_2}&\le \epsilon_i \label{eq:Delta2}
\end{align}
\end{subequations}
for all $i\in[p]$, and let $\Delta$ be such that $\Delta > \Delta_1 +\Delta_2$.
Assume that $\left| U(t) \right| \ge p-q$ for all $t \ge 0$ where
\begin{equation}\label{eq:attackRelax}
U(t) := \bigcap_{\max\{t-\Delta,0\} \le\tau\le t}\{i\in[p]: a_i(\tau) = 0\}.
\end{equation}
\end{asm}

Under this relaxed assumption, at any time $t$, there are at least $p-q$ sensors that are attack free for the last $\Delta$ seconds.
As will be shown in Theorem~\ref{thm:signalRecovery},
it ensures the existence of an ``attack free'' index set $I^*(t)\in\binom{[p]}{p-q}$ for each $t$ such that all estimates $\hat z_i(t)$'s with respect to $i\in I^*(t)$ obeys Lemma \ref{lem:gauthierObserver} so that \eqref{eq:errordetection} is violated with $I = I^*(t)$.
Now, one idea to estimate the state $x(t)$ under $q$-sparse sensor attack is to prepare all $\binom{p}{p-q}$ index sets $I \in \binom{[p]}{p-q}$, and test the attack detection \eqref{eq:errordetection} for all of them at each time instant\footnote{In fact, it is checked at each sampling instant since the proposed scheme is implemented in a digital computer.}.
Then, one can always find at least one set $I^*$ that violates \eqref{eq:errordetection} implying that there is no influential attack in $\{\hat z_i\}_{i \in I^*}$.
Therefore, the true state $x(t)$ is estimated by $\hat x(t) = \Psi^{I^*}(\hat z_{I^*}(t))$ with the estimation error discussed in Theorem \ref{thm:staticDetection}.
However, testing \eqref{eq:errordetection} at each sampling instant with all index sets is computationally heavy.
This burden may be relieved by introducing a simple switching algorithm as in the following theorem. 

\begin{thm}\label{thm:signalRecovery}
Under Assumptions \ref{asm:uniformObservableDecomposition} and \ref{asm:sparsity2}, assume that system \eqref{eq:system} is $2q$-redundant observable.
Let $\Lambda : [ \binom{p}{p-q} ] \ra \binom{[p]}{p-q}$ be a bijection set-valued map, such that $\{ \Lambda(i) : i \in \binom{p}{p-q} \} = \binom{[p]}{p-q}$.
Consider a switching signal $\sigma(t)$ generated from $\sigma(0)=1$ by the update rule
\begin{equation}\label{eq:updateRule}
\sigma(t^+) \leftarrow \left( \sigma(t) \; {\rm mod} \; \binom{p}{p-q} \right) + 1
\end{equation}
whenever
\begin{align}\label{eq:thm}
	\begin{split}
	& \| \hat z_{\Lambda(\sigma(t))}(t) - \Phi_{\Lambda(\sigma(t))}(\Psi^{\Lambda(\sigma(t))}(\hat z_{\Lambda(\sigma(t))}(t))) \|_\infty \\
	&\quad > \overline{\mathrm{Lip}}(\id_{\R^{\NN_{\Lambda(\sigma(t))}}} - \Phi_{\Lambda(\sigma(t))}\circ\Psi^{\Lambda(\sigma(t))}) \delta(t).
	\end{split}
\end{align} 
Then, the state estimate for $x(t)$ is given by
$$\hat x(t) = \Psi^{\Lambda(\sigma(t))} (\hat z_{\Lambda(\sigma(t))}(t))$$
which has the property
\begin{equation}\label{eq:conclusion}
	\|\hat x(t) - x(t) \|_\infty \le  \frac{\max_{I\in\binom{[p]}{p-q}} \overline{\mathrm{Lip}}(\id_{\R^{\NN_{I}}} - \Phi_I\circ\Psi^I) +1 }{\min_{J \in \binom{[p]}{p-2q}} \underline{\mathrm{Lip}}(\Phi_J)} \delta(t)
\end{equation}
for all $t \ge 0$ except at the switching times of $\sigma(t)$.
\end{thm}

Update of the switching signal $\sigma$ in Theorem \ref{thm:signalRecovery} is understood as follows.
Whenever the value of $\sigma(t)$ is updated at time $t$, condition \eqref{eq:thm} is checked again at the same time $t$ with the updated $\sigma(t^+)$ until the inequality is violated (i.e., consecutive updates can occur). 
This update does not repeat for infinitely many times as is shown in the proof.
While this behavior can be described more rigorously by introducing a hybrid time domain $(t,j)$ with continuous time $t$ and jump time $j$, as done in, e.g., \cite{Goebel09}, we do not follow such convention for the sake of simplicity.

{\em Proof:}
Consider a sequence of time interval $\TT_0 := [0, \Delta]$ and $\TT_k := (t_{k-1},t_k]$, for $k=1,2,\cdots$, where $t_k := \Delta + k(\Delta - \Delta_1 - \Delta_2)$.
Then, $\cup_{k=0}^\infty \TT_k = \{t : t \ge 0\}$.
We claim that, for each $\TT_k$, $k=0, 1, \cdots$, there is a natural number $m_k \le \binom{p}{p-q}$ such that $\Lambda(m_k) \subseteq U(t_k)$ and
\begin{multline}\label{eq:1}
\|\hat z_{\Lambda(m_k)}(\tau) - \Phi_{\Lambda(m_k)}(x(\tau)) \|_\infty \le \delta(\tau), \;\; \forall \tau \in \TT_k. 
\end{multline}
Indeed, when $k=0$, the claim follows from Assumption \ref{asm:sparsity2} and Lemma \ref{lem:gauthierObserver}.
For the case when $k \ge 1$, Assumption \ref{asm:sparsity2} ensures the existence of $m_k$ such that $\Lambda(m_k) \subseteq U(t_k)$, implying that $a_i(\tau)=0$ for $i \in \Lambda(m_k)$ and $\tau \in (t_k-\Delta,t_k]$.
Then, for the state $\hat z_i$ where $i \in \Lambda(m_k)$, we observe the following.
First, since the reset rule \eqref{eq:resetRule} guarantees that $\|\hat z_i(t)\|_\infty \le \max\{ 2\eta_iM_{z,i}, \epsilon_i \}+M_{z,i}$ for all $t \ge 0$, if no reset occurs for $\hat z_i$ during $(t_k-\Delta,t_k-\Delta+\Delta_1]$, then we have that $\|\hat z_i(\tau) - z_i(\tau)\|_\infty \le \epsilon_i$ for $\tau \in [t_k-\Delta+\Delta_1,t_k]$ by \eqref{eq:Delta1} and Lemma \ref{lem:gauthierObserver} because $\|\hat z_i(t-\Delta) - z_i(t-\Delta)\|_\infty \le \|\hat z_i(t-\Delta)\|_\infty + \|z_i(t-\Delta)\|_\infty \le \max\{ 2\eta_i M_{z,i}, \epsilon_i\} + 2 M_{z,i}$.
If not, that is, a reset \eqref{eq:resetRule} occurs at $\bar t \in (t_k-\Delta,t_k-\Delta+\Delta_1]$, then it follows from \eqref{eq:Delta2} and Lemma \ref{lem:gauthierObserver} that $\|\hat z_i(\tau) - z_i(\tau)\|_\infty \le \epsilon_i$ for $\tau \in (\bar t+\Delta_2,t_k]$.
In both cases, it can be seen that \eqref{eq:1} holds because $\epsilon_i \le \delta(\tau), \forall i\in [p]$.

Now, it is seen that \eqref{eq:1} implies that $\| \hat z_{\Lambda(m_k)}(t) - \Phi_{\Lambda(m_k)} ( \Psi^{\Lambda(m_k)}(\hat{z}_{\Lambda(m_k)}(t)) )\|_\infty \le \overline{\mathrm{Lip}} (\id_{\R^{\NN_{\Lambda(m_k)}}} - \Phi_{\Lambda(m_k)} \circ \Psi^{\Lambda(m_k)}) \delta(t)$ for every $t \in \TT_k$, $k=0, 1, \cdots$.
Therefore, according to the update rule of $\sigma$, there is a maximum of $\binom{p}{p-q}-1$ (consecutive) switchings of $\sigma(t)$ (until \eqref{eq:thm} is violated) in each time interval $\TT_k$, $k=0, 1, \cdots$.
(However, $\sigma$ does not necessarily become identical to $m_k$.)
Then, the proof is completed from the upper bound of the estimation error in Theorem \ref{thm:staticDetection}, by considering that the set $I \in \binom{[p]}{p-q}$ is arbitrary.
\hfill$\blacksquare$

\begin{rem}
	Compared to the result in \cite{Chanhwa15ECC}, where the search for uncompromised sensors (or, search for suitable $\sigma$) is performed at each sampling time, Theorem \ref{thm:signalRecovery} only searches when the inequality condition of \eqref{eq:thm} is satisfied. 
	In this sense, the computational burden is relieved in the proposed scheme.
\end{rem}

Note that no zeno behavior appears in the switching scheme of Theorem \ref{thm:signalRecovery} because no infinitely many switchings occur in any finite time interval.
On the other hand, since the estimator is implemented on a digital computer in practice, a few sampling delays may occur owing to consecutive updates, and during these delays, the state estimation can be corrupted, which is seen in the simulation results in the next section.

\section{Simulation Results}\label{sec:example}

We consider a numerical example of system \eqref{eq:system} given as
\begin{align*}
\begin{bmatrix} \dot x_1 \\ \dot x_2 \\ \dot x_3 \end{bmatrix}
&= \begin{bmatrix} -2x_1 - x_2^3  \\ -x_2 \\ -x_2\cos x_2 + \sin x_2 - x_3  \end{bmatrix}
+ \begin{bmatrix} 1+3x_2^2 \\ 1 \\ \cos x_2 \end{bmatrix} u \\
\begin{bmatrix}y_1\\y_2\\y_3\\y_4 \end{bmatrix}
&=\begin{bmatrix}
x_1 + x_2 - x_2^3 - \sin x_2 + x_3\\
x_1 + \sin x_2 - x_2^3 - x_3\\
-x_1 + x_2^3 + x_2\\
-x_2 -\sin x_2 + x_3
\end{bmatrix}
+ \begin{bmatrix}
a_1 \\ a_2 \\ a_3 \\ a_4
\end{bmatrix} + \begin{bmatrix} v_1 \\ v_2 \\ v_3 \\ v_4 \end{bmatrix}
\end{align*}
where $u(t)=0.25\sin(0.2\pi t)-0.1$, for which it is verified that the state $x$ remains in $\XX=\{x\in\R^3:\|x\|_\infty \le 0.5 \}$ with sufficiently small initial conditions.
The bounded measurement noises $\{v_i\}_{i=1}^4$ are generated from a uniform distribution over $[-10^{-6},10^{-6}]$.
For this system, the function $\Phi:\XX\ra\R^{\NN}$ in \eqref{eq:observablePartMapping} and \eqref{eq:Phi} is computed by
\begin{equation*}
\begin{bmatrix}
\Phi_1(x)\\\Phi_2(x)\\\Phi_3(x)\\\Phi_4(x)
\end{bmatrix}
\!=\!
\begin{bmatrix}
h_1(x)\\L_f h_1(x)
\\
h_2(x)\\L_f h_2(x)
\\
h_3(x)\\L_f h_3(x)
\\
h_4(x)
\end{bmatrix}
=
\begin{bmatrix}
x_1 + x_2 - x_2^3 - \sin x_2 + x_3\\
\!-\!2x_1 \!+\! \sin x_2 \!-\! x_2 \!+\! 2 x_2^3 \!-\!x_3\\
x_1 + \sin x_2 - x_2^3 - x_3\\
-2 x_1 - \sin x_2 + 2 x_2^3 + x_3\\
-x_1 + x_2^3 + x_2\\
2 x_1 - x_2 - 2 x_2^3\\
-x_2 -\sin x_2 + x_3
\end{bmatrix}
\end{equation*}
with $\NN = (2,2,2,1)$.
It can be seen that each $\Phi_i$ transforms the system into a uniformly observable subsystem with respect to $y_i$,
and the stack of all observable parts $z$ remains in the set ${\mathcal Z} := \{z\in \R^7:\|z\|_\infty \le 2 \}$.
One can also ensure that the above system is $2$-redundant observable by verifying that $\Phi_J$ is an injective immersion for every $J\in\binom{[4]}{2}$.

\begin{figure}[!t]
	\centering
	\begin{subfigmatrix}{2}
		\subfigure[Injected sensor attack signal $a_2(t)$ (solid) and $a_3(t)$ (dash-dot)]{
			\includegraphics[width=.225\textwidth]{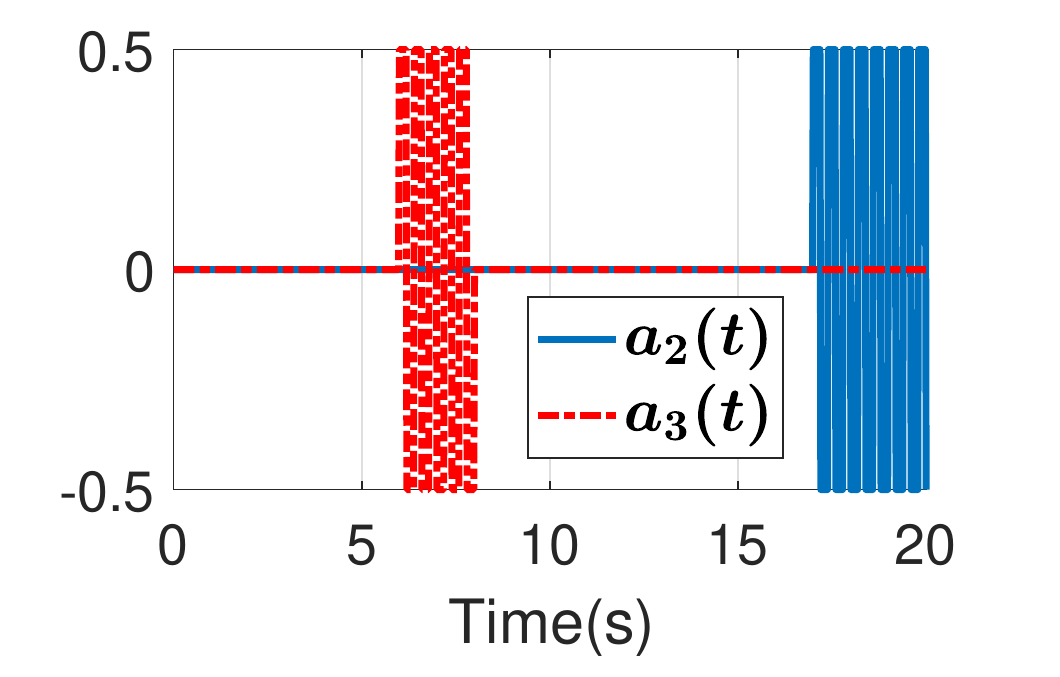}
			\label{fig1:subfig1}}
		\vrule
		\subfigure[Index function $\sigma(t)$]{
			\includegraphics[width=.225\textwidth]{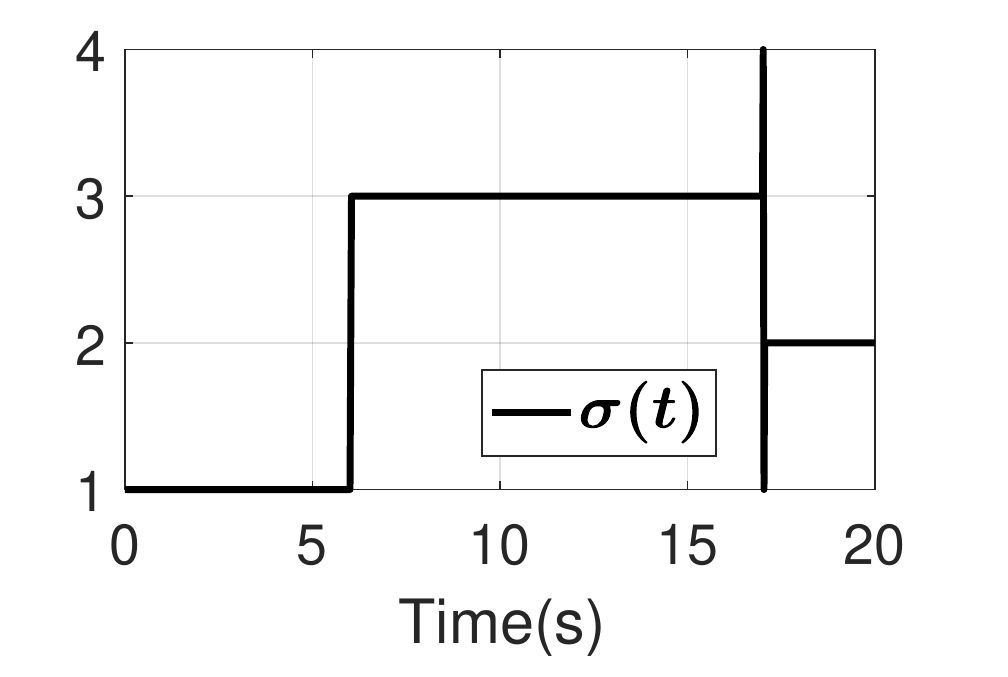}
			\label{fig1:subfig4}}
	\subfigure[Estimation error $\hat x(t) - x(t) = \Psi^{\Lambda(\sigma(t))}(\hat z_{\Lambda(\sigma(t))}(t)) - x(t) $]{
		\includegraphics[width=.225\textwidth]{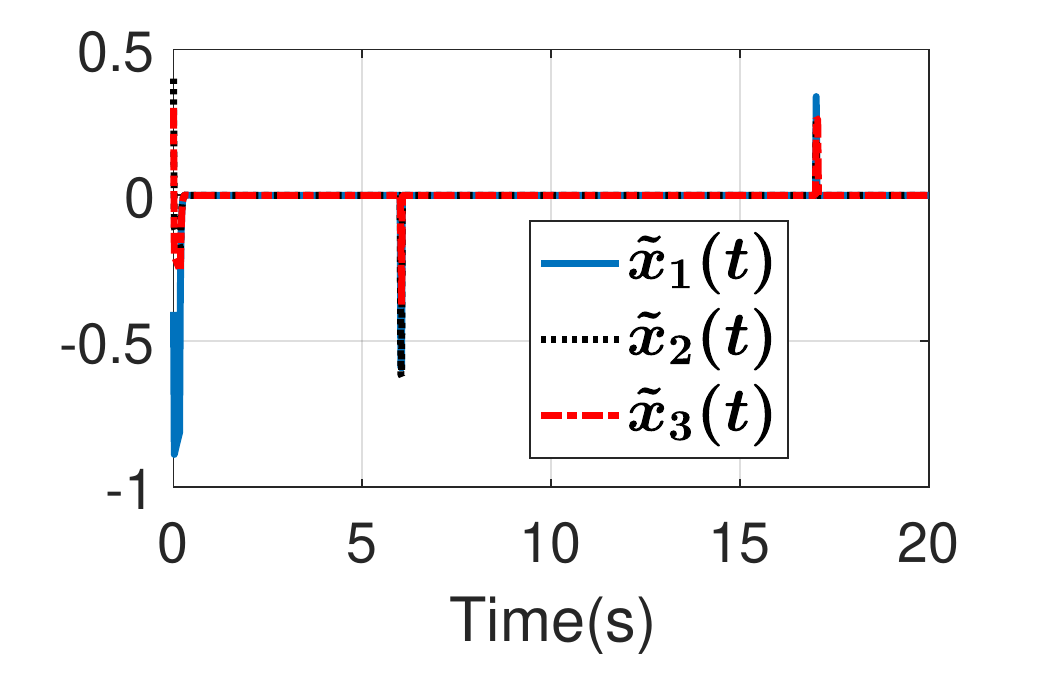}
		\label{fig1:subfig3}}
	\vrule
	\subfigure[Estimation error $\Psi^{\Lambda(3)}(\hat z_{\Lambda(3)}(t)) - x(t)$]{
		\includegraphics[width=.225\textwidth]{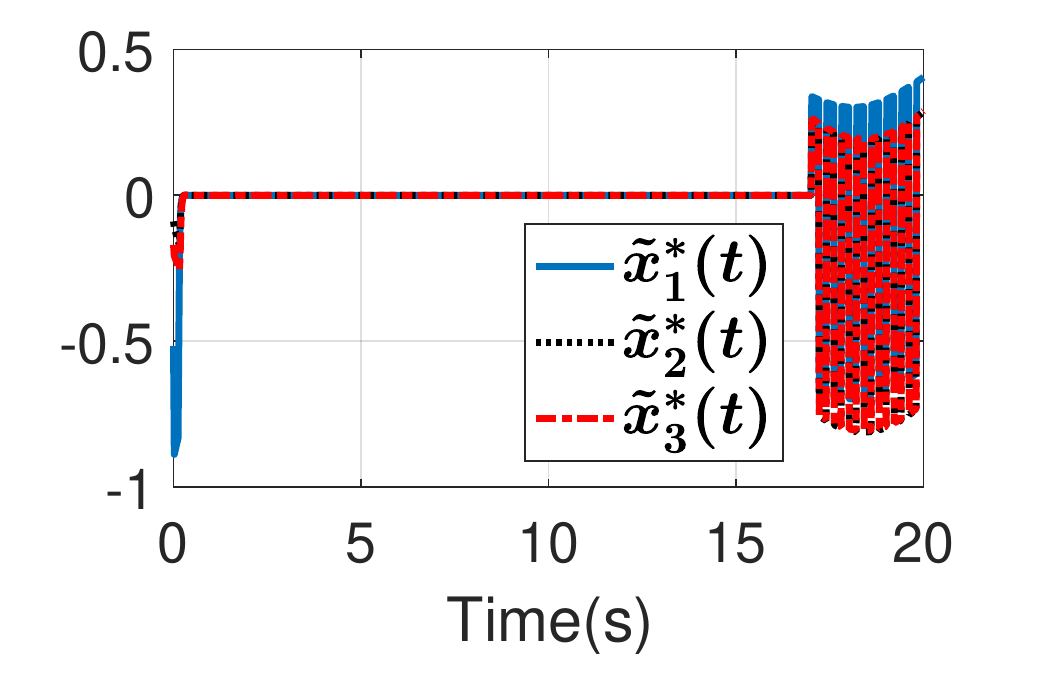}
		\label{fig1:subfig2}}
\end{subfigmatrix}
	\caption{Simulation results
	}
\label{fig1}
\end{figure}

As the system is $2$-redundant observable, resilient state estimation is possible under up to 1-sparse attack.
Therefore, let us suppose the attack scenario depicted in Fig.~\ref{fig1:subfig1}; a square wave $a_3$ is injected into the third sensor on time interval $t\in[6,8]$; then, the second sensor is attacked by a similar square wave $a_2$ on $t\in[17,20]$.
$a_1$ and $a_4$ remain zero.
This scenario satisfies Assumption \ref{asm:sparsity2} with $\Delta = 8.37$; it is 1-sparse for each time $t$ and the change in attacked sensor occurs intermittently.

Partial observers for individual uniformly observable subsystems are designed with $\theta_i = 32$ for $i\in [4]$, which yields
$\delta(t)=\max\{671\times\exp(-4t),4.74\times 10^{-4}\}$.

For the recovery of state $x$, we choose four left inverse functions $\Phi_{\Lambda(i)}^{-1}$ for each $i=1,\cdots,4$, where $\Lambda(i)=[4]-\{i\}$, as in Table \ref{tab:Phi}.
With these functions, as in \eqref{eq:leftinverse} and \eqref{eq:leftinverseReplaced}, the Lipschitz-extended left inverse of $\Phi_{\Lambda(i)}$ is obtained by
\begin{align*}
\Psi^{\Lambda(i)}: \R^{\NN_{\Lambda(i)}}&\ra \XX\\
\zz_{\Lambda(i)} &\mapsto \sat(\Phi_{\Lambda(i)}^{-1}(\sat(\zz_{\Lambda(i)},\,2)) ,\,0.5)
\end{align*}
for each $i$.
It is noted that the Lipschitz constant of $\Psi^{\Lambda(i)}(\cdot)$ on $\R^{\NN_{\Lambda(i)}}$ is less than or equal to the Lipschitz constant of $\Phi_{\Lambda(i)}^{-1}(\cdot)$ on ${\mathcal Z}^{\Lambda(i)} = \{z_{\Lambda(i)}\in\R^{\NN_{\Lambda(i)}}: \|z_{\Lambda(i)}\|_\infty\le 2\}$ owing to the two saturation functions, and the Lipschitz constant of $\Phi$ is greater than or equal to the Lipschitz constant of $\Phi_{\Lambda(i)}$.
Hence, for simplicity, we take a conservative bound for the right hand side of the condition \eqref{eq:thm} as
\begin{align*}
&\overline{\mathrm{Lip}}(\id_{\R^{\NN_{\Lambda(\sigma)}}} - \Phi_{\Lambda(\sigma)}\circ\Psi^{\Lambda(\sigma)})\\
&\le 1+ \overline{\mathrm{Lip}}(\Phi)\times\max_{i\in[4]}\left\{\overline{\mathrm{Lip}}(\Phi^{-1}_{\Lambda(i)}|_{\mathcal{Z}^{\Lambda(i)}} )\right\}
\le 1+ 7\times 770.
\end{align*}
By this simplification, the upper bound of the estimation error in Theorem \ref{thm:signalRecovery} is increased (as the numerator in the upper bound is replaced by $1+7 \times 770$); however, the simulations show that this is not a large sacrifice.

\begin{table}[!tb]
	\caption{Left inverse functions of $\Phi_{\Lambda(i)}$, ~$i =1,2,3,4$}
	\label{tab:Phi}
 {\small
		\begin{center}
			\begin{tabular}{c|c}
				\toprule
				\\\addlinespace[-1.5ex]
				$\begin{matrix}\Lambda(1)=\{2,3,4\}\\\Phi_{\{2,3,4\}}^{-1}(\zz):\R^5\ra\R^3\end{matrix}$ &
				$\begin{bmatrix}
				\zz_3+\zz_4 + (2\zz_3 + \zz_4)^3\\
				2\zz_3 + \zz_4\\
				-2\zz_1 - \zz_2 + \sin(2\zz_3 + \zz_4)
				\end{bmatrix}$
				\\
				\\\addlinespace[-1.5ex]
				\hline
				\\\addlinespace[-1.5ex]
				$\begin{matrix}\Lambda(2)=\{1,3,4\}\\\Phi_{\{1,3,4\}}^{-1}(\zz):\R^5\ra\R^3\end{matrix}$	& 
				$\begin{bmatrix}
				-\zz_1 - \zz_2 + (2\zz_3 + \zz_4)^3\\
				2\zz_3 + \zz_4\\
				\frac{1}{2}(2\zz_1 + \zz_2 + \zz_5) + \sin(2\zz_3 + \zz_4)
				\end{bmatrix}$
				\\ 
				\\\addlinespace[-1.5ex]
				\hline
				\\\addlinespace[-1.5ex]
				$\begin{matrix}\Lambda(3)=\{1,2,4\}\\\Phi_{\{1,2,4\}}^{-1}(\zz):\R^5\ra\R^3\end{matrix}$ &	
				$\begin{bmatrix}
				-\zz_1  - \zz_2 - (2\zz_3 +\zz_4 + \zz_5)^3\\
				-2\zz_3 - \zz_4 -\zz_5\\
				-2\zz_3 - \zz_4 + -\sin(2\zz_3 + \zz_4 + \zz_5)
				\end{bmatrix}$
				\\ 
				\\\addlinespace[-1.5ex]
				\hline 
				\\\addlinespace[-1.5ex]
				$\begin{matrix}\Lambda(4)=\{1,2,3\}\\\Phi_{\{1,2,3\}}^{-1}(\zz):\R^6\ra\R^3\end{matrix}$	&  
				$\begin{bmatrix}
				-\zz_1 -\zz_2 + (2\zz_5 + \zz_6)^3\\
				2\zz_5 + \zz_6\\
				-2\zz_3 - \zz_4 + \sin ( 2\zz_5 + \zz_6)
				\end{bmatrix}$
				\\\addlinespace[-1.5ex]  \\ \bottomrule 
			\end{tabular}
		\end{center}	
	}
\end{table}

The detection algorithm and estimator are simulated in discrete-time with a sampling time of $0.02{\mathrm{s}}$.
Fig.~\ref{fig1} summarizes the outcome.
Fig.~\ref{fig1:subfig4} shows that as $\sigma(t)$ begins from $1$, it jumps to $2$ and to $3$ consecutively, immediately after the attack signal is injected in the measurement.
Note that $\sigma(t)=3$ is the correct index because $\Lambda(3)=\{1,2,4\}$ does not contain the index of compromised output $y_3$.
As the jumps of $\sigma(t)$ to $2$ and $3$ take slightly more time in practice, the estimation errors become large during these short periods as seen in Fig.~\ref{fig1:subfig3}.
Similar behavior is observed when the attack is injected into the output $y_2$ at $t=17$.
It is seen that $\sigma(t)$ jumps as $3 \to 4 \to 1 \to 2$ around this time, and the estimate $\hat x(t)$ correctly recovers the state $x(t)$ owing to the switching of $\sigma(t)$.
Without this switching, an accurate estimation for $x(t)$ is not obtained, which can be seen, for example, by looking at the signal $\Psi^{\Lambda(3)}(\hat z_{\Lambda(3)}(t)) - x(t)$ for all time frames in Fig.~\ref{fig1:subfig2}.
During the sensor attack on $y_2$, the estimate $\Psi^{\Lambda(3)}(\hat z_{\Lambda(3)}(t))$ cannot recover the state although it can for the attack on $y_3$.

\section{Conclusion}\label{sec:con}

In this paper, we proposed a solution to a resilient state estimation problem for uniformly observable nonlinear systems with redundant sensors.
A switching algorithm that makes use of the detection algorithm of sensor attacks is designed to search for a combination of uncompromised sensors successfully and generate accurate estimates that are insensitive to sparse malicious attacks.
Lastly, an illustrative example and its simulation results, which demonstrate the effectiveness of the proposed estimator, are provided.

\appendix

\noindent{\it Proof of Lemma \ref{lem:gauthierObserver}:}
It is a slight modification of the proof of \cite[Lemma 3.2.2]{shim2000passivity}.
Let $\xi_i := \mathrm{diag}(1,1/\theta_i,\cdots,1/\theta_i^{n_i -1})(\hat z_i - z_i)$ and $\tilde P_i\in\R^{n_i \times n_i}$ be the unique positive definite solution of $ 0 = -\tilde P_i - \tilde P_i A_i - A_i^T \tilde P_i + C_i^T C_i$.
Then, as in \cite{shim2000passivity}, it yields
\begin{align*}
	\frac{d}{dt}(\xi_i^T \tilde P_i \xi_i) &\le -\frac{\theta_i}{2}\xi_i^T \tilde P_i \xi_i + 2\theta_i v_i C_i \xi_i\\
	&\le -\frac{\theta_i}{2}\xi_i^T \tilde P_i \xi_i + \frac{2\theta_i M_{v,i}}{\sqrt{\lambda_{1,i}}} \sqrt{\xi_i^T \tilde P_i \xi_i}\\
	&\le -\frac{\theta_i}{4}\xi_i^T \tilde P_i \xi_i + 4\theta_i \frac{M_{v,i}^2}{\lambda_{1,i}},
\end{align*}
where $\lambda_{1,i}$ is the smallest eigenvalue of $\tilde P_i$, and it follows that
\begin{align*}
	\|\xi_i(t) \|_\infty^2
	&\le\frac{\xi_i^T(t) \tilde P_i \xi_i(t)}{\lambda_{1,i}}
	\le\frac{e^{-\frac{\theta_i}{4}t}\xi_i^T(0)\tilde P_i \xi_i(0) }{\lambda_{1,i}}+ \frac{16M_{v,i}^2}{\lambda_{1,i}^2}\\
	&\le 2\max\left\{\frac{\lambda_{2,i}n_i e^{-\frac{\theta_i}{4}t}\|\xi_i(0) \|_\infty^2}{\lambda_{1,i}}, \frac{16M_{v,i}^2}{\lambda_{1,i}^2} \right\},
\end{align*}
in which $\lambda_{2,i}$ is the largest eigenvalue of $\tilde P_i$. Finally, we obtain \eqref{eq:partialObsConvergence} with
\begin{equation*}
	\eta_i(\theta_i) := \sqrt{\frac{2n_i \lambda_{2,i}}{\lambda_{1,i}}}\theta_i^{n_i-1},\quad \epsilon_i(\theta_i):=\frac{4\sqrt{2}M_{v,i}}{\lambda_{1,i}}\theta_i^{n_i-1}
\end{equation*}
from the fact that $\|\xi_i \|_\infty\le \|\hat z_i - z_i \|_\infty\le \theta_i^{n_i-1}\|\xi_i \|_\infty$.

\noindent{\it Proof of Proposition \ref{prop:locallyBi-Lipschitz}:}
Note that $\phi$ is Lipschitz on $X$ because it is continuously differentiable. Thus, it is sufficient to show $$\inf\limits_{\substack{x\neq x'\\x,x'\in X}}\frac{\|\phi(x)-\phi(x') \|_\infty}{\|x-x' \|_\infty}>0.$$
Suppose, for the sake of contradiction, there exist sequences $\{x_{i}\}_{i=1}^\infty $ and $\{x_{i}'\}_{i=1}^\infty$ in $X$ such that $x_i \not = x_i'$ and 
\begin{equation}\label{eq:supposeNotLipschitz}
\lim\limits_{i\ra\infty}\frac{\|\phi(x_{i})-\phi(x_{i}') \|_\infty}{\|x_{i}-x_{i}' \|_\infty}=0.
\end{equation}
By the Bolzano-Weierstrass theorem, without loss of generality (by taking any convergent subsequence if necessary), we may assume that $\{x_{i}\}_{i=1}^\infty$ and $\{x_{i}'\}_{i=1}^\infty $ converge to points $x_{\infty}$ and $x_{\infty}'$ in $X$, respectively.
If $x_{\infty}\neq x_{\infty}'$, then $\phi(x_{\infty})=\phi(x_{\infty}')$, which contradicts the injectivity of $\phi$.
If $x_{\infty}= x_{\infty}'$,
by continuous differentiability of $\phi$,
it is derived that
\begin{equation}\label{eq:contiDiff}
\lim\limits_{i\ra\infty}\frac{\| \phi(x_{i})-\phi(x_{i}') -   D\phi ({x_{\infty}}) \cdot  (x_{i}-x_{i}')    \|_\infty}{\|x_{i}-x_{i}' \|_\infty}=0,
\end{equation}
where $D\phi (x_{\infty})$ denotes the Jacobian matrix of $\phi$ at ${x_{\infty}}$.
Hence, it follows from the combination of \eqref{eq:contiDiff} and \eqref{eq:supposeNotLipschitz} that
$$\lim\limits_{i\ra\infty}\frac{\|   D\phi ({x_{\infty}}) \cdot  (x_{i}-x_{i}')    \|_\infty}{\|x_{i}-x_{i}' \|_\infty}=0,$$
which contradicts the fact that $D\phi (x_\infty)$ has full column rank.

\end{document}